\documentclass{article}
\usepackage{spconf,amsmath,graphicx}
\usepackage{amssymb}
\usepackage[utf8]{inputenc}
\usepackage{bm}
\usepackage{xcolor}
\usepackage{cite}
\newtheorem{assumption}{A.}
\usepackage{booktabs}
\usepackage{psfrag}
\usepackage{color}
 \allowdisplaybreaks

\def\Vbo{\Vb^{(1)}}
\def\Vbt{\Vb^{(2)}}

\def\vb{\mathbf v}

\def\Vb{\mathbf V}

\def\thetab{\mathbf \theta}

\def \rhob{\pmb \rho}

\title{Learning Connectivity and Higher-order Interactions\\ in Radial Distribution Grids
}

 \name{Qiuling Yang$^{\dagger}$, Mario Coutino$^{\ddagger}$, Gang Wang$^{\ast}$, Georgios B. Giannakis$^{\ast}$, and Geert Leus$^\ddagger$}
 
\address{$^{\dagger}$ School of Automation, Beijing Institute of Technology
\\$^{\ddagger}$ Dept. of Microelectronics, Delft University of Technology
\\$^{\ast}$  Dept. of ECE and Digital Technology Center, University of Minnesota
}

\begin{document}
\ninept
\maketitle
\begin{abstract}

To perform any meaningful optimization task, distribution grid operators need to know the topology of their grids. 
Although power grid topology identification and verification has been recently studied, 
discovering instantaneous interplay among subsets of buses, also known as higher-order interactions in recent literature, has not yet been addressed. 
The system operator can benefit from having this knowledge when re-configuring the grid in real time, to minimize power losses, balance loads, alleviate faults, or for scheduled maintenance.  
Establishing a connection between the celebrated exact distribution flow equations
and the so-called self-driven graph Volterra model, this paper puts forth a nonlinear topology identification algorithm, that is able to reveal both the edge connections as well as their higher-order interactions. Preliminary numerical tests using real data on a 47-bus distribution grid showcase the merits of the proposed scheme relative to existing alternatives.

\end{abstract}
%
%

\section{Introduction}
\label{sec:intro}
Full awareness of the distribution grid topology is required for a system operator to perform any tasks related to monitoring, control, optimization, and planning \cite{golsha2014tsg}, \cite{fitee2019wgcs}. For instance, the increasing penetration of distributed renewable generation in power grids nowadays can cause sizable and frequent voltage fluctuations, as well as power line congestion. In this case, the distribution grid can be reconfigured or planned to alleviate excessive voltage drops, reduce line congestion, or minimize power losses. Nonetheless, taking these measures entails full knowledge of the topology and the interactions between groups of nodal buses.

High-voltage power transmission networks are typically operated under fixed topology (i.e., there is no frequent network reconfiguration during operation), and their topology capturing how buses are connected with each other through electric lines is typically available to system operators \cite{save2013cdc}. However, in low- and medium-voltage residential distribution grids, due to the increasing deployment of information and communication technologies, which retrofit the existing power infrastructure by installing new devices, real-time topology information is in general not available \cite{save2013cdc}. In this context, distribution grid topology identification is a prerequisite for subsequent monitoring, control, and optimization tasks.

A number of methods have been proposed to identify the connections present in a power network; see, for example, \cite{gg2018proceeding} for a recent overview. The grid topology was reconstructed by means of impedance estimation
at the point of connection of buses in \cite{arxiv2018low}. 
 Nevertheless, since multiple topologies can report similar impedance, 
 the success of estimating the correct topology cannot be guaranteed in general. 
Using a linear approximation model, topology processing was performed by perturbing power injections at certain buses and observing nodal voltage responses \cite{probing}.
A data-driven
 topology identification algorithm was advocated by using the signs of the elements in the inverse sample covariance matrix of nodal voltage magnitudes in \cite{save2013cdc}. 
 Even though these methods improve with respect to the impedance-based methods, they do not account for the intrinsic nonlinear dependencies between nodal voltage measurements, yielding sub-optimal performance. Finally, leveraging the DC power flow model, a blind topology identification algorithm based on power injections was developed for transmission networks in~\cite{li2013blind}. 

To capture the nonlinear connectivity and dynamics in the data, topology identification methods combining partial correlations and kernels have
recently been investigated
\cite{liang2017topology}, \cite{gg2018proceeding}, \cite{coutino2019state}.  While it is possible to adopt these methods to identify the distribution grid topology, they face two challenges in practice: First, selecting proper kernels requires cross-validations, or solving computationally
involved optimization tasks.
Second, kernels used to model the nonlinear interactions between the data do not allow interpreting the interplay that different buses in children branches exhibit. 
Although such nonlinear topology identification methods can learn meaningful connections, they lack the ability to unravel interactions occurring among a group of buses, namely higher-order interactions. This limitation is mostly due to their reliance on the celebrated structural equation models (SEMs)~\cite{SEM},  \cite{coutino2019holink}. Indeed, SEMs have been successfully used to identify network topology in diverse applications \cite{gg2018proceeding}. However, since they build on pair-wise interactions, they cannot capture higher-order interactions among a group of buses.   

Building on recent advances in understanding higher-order nodal interactions over  graphs~\cite{benson2016higher},
we put forward a principled approach to unveiling, not only the edge connectivity but also the higher-order interactions in a distribution grid. Our approach is motivated by the nonlinear distribution grid flow model as well as the recent self-driven graph Volterra models \cite{coutino2019holink}. 
Specifically, focusing on nodal voltage magnitude time-series, 
we establish a nonlinear model for voltage measurements that explores  higher-order interactions inherited by parent-children bus relations. Drawing a connection with the self-driven graph Volterra models, offers a framework for simultaneous topology and higher-order interaction identification. Preliminary tests are provided to showcase the practical merits of our proposed approach.
\begin{figure}[t]
\centering
\includegraphics[width=0.85\linewidth]{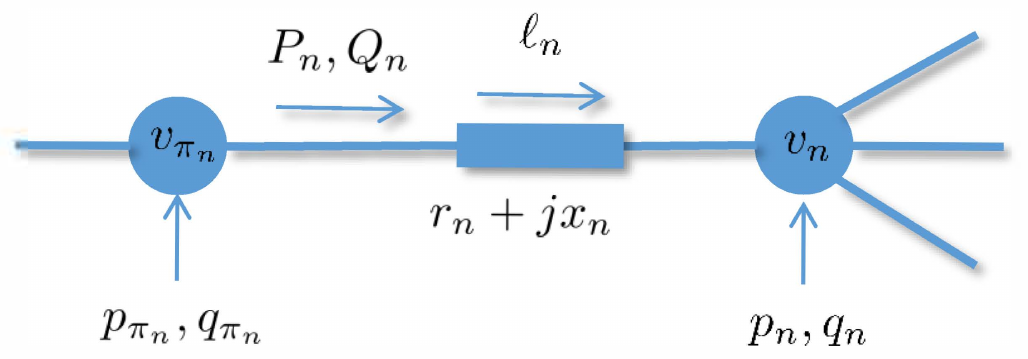}
\caption{Bus $n$ is connected to its unique parent $\pi_n$ via line $(\pi_n,n)$.}
\label{fig:powerflow}
\end{figure}

\section{Distribution Grid Modelling}
\label{sec:syst}
\def\cN{\mathcal{N}}
\def\oN{\cN_0}
\def\cE{\mathcal{E}}
\def\cG{\mathcal{G}}
\def\bB{\mathbf{B}}
\def\bb{\mathbf{b}}

Consider a radial distribution grid represented by a graph 
$\mathcal{G}:=(\oN,\cE)$, where the vertex set $\oN$ collects the indices of the nodal buses, and the edge set $\cE$ the power distribution lines. Here, one can write $\oN := \{0,\cN\}$ with index $0$ denoting the \emph{root} bus (i.e., substation),  and all \emph{non-root} buses collected in $\cN$. It is evident for a tree network that the number of non-root buses and power lines is the same, i.e., $|\cN| =|\cE|= N$.

Considering the radial nature of distribution grids, that is $\cG$ exhibits a tree structure, every non-root bus $n \in \mathcal N$ has a unique parent bus, which is denoted by $\pi_n$. The $n$-th power line $(\pi_n, n) \in \mathcal E $, which connects bus $n$ and its parent bus $\pi_n$, is modeled by its impedance $z_n = r_n+j x_n$. Moreover, the squared current magnitude of line $n$ is denoted by $\ell_n$, and the complex power flowing from bus $\pi_n$ to bus $n$ is denoted by $S_n = P_n+j Q_n$. See Fig. \ref{fig:powerflow} for an illustration.
 
To make explicit the connectivity of the network $\mathcal G $, let us define  the so-called bus-branch incidence matrix $\bB \in \mathbb {R}^{N \times(N+1)}$ with elements  
\begin{equation}\notag
    B_{i,j}=\left\{\begin{array}{cl}{-1,} & {\text { if } i \in \mathcal C_j} \\ 
    {1,} & {\text { if } j \in \mathcal C_i} \\ 
    {0,} & {\text { otherwise }}\end{array}\right.
\end{equation}
where $\mathcal C_j\subseteq \mathcal N $ denotes the set of children nodes for bus $j$.
Partitioning $\bB$ into the first and the rest of its columns gives rise to
\begin{equation}\notag
    \bB:=\left[\begin{matrix}\bb_{0}~~ \tilde{\bB} \end{matrix}\right]
\end{equation}
where $\tilde{\mathbf{B}}\in\{-1,0,1\}^{N\times N}$ is the so-called reduced branch-bus incidence matrix \cite{liang2015smartgrid}. Evidently,   $\tilde{\bB}$ is a square matrix due to the radial configuration of the network.

Using the branch flow model~\cite{baran1989optimal}, the power flow over lines can be modeled by the following equations
\begin{subequations}\label{eq:powerflowequations}
    \begin{align}
     s_{n} &=\sum_{i \in \mathcal{C}_{n}} S_{i}-S_{n}+\ell_{n} z_{n}, \quad \qquad\quad ~\;\forall n\in\mathcal N \label{eq:sn}\\ 
     v_{n} &=v_{\pi_{n}}-2 \operatorname{Re}\left[z_{n}^{\star} S_{n}\right]+\ell_{n}\left|z_{n}\right|^{2}, \quad  \forall n\in\mathcal N \label{eq:vn}\\
     \left|S_{n}\right|^{2} &=v_{\pi_{n}} \ell_{n}, \qquad\qquad\qquad\qquad\qquad\quad  \forall n\in\mathcal N \label{eq:ssn}
    \end{align}
\end{subequations}
where $v_n$ denotes the squared voltage magnitude of bus $n \in \mathcal N_0 $, and $s_n = p_n + j q_n$ the complex power injected into bus $n$. 
As we focus on a grid-connected distribution feeder, the measurements at the substation ($n=0$) are assumed to be constant. 
For future reference, we collect all non-root buses and line quantities into $N$-dimensional column vectors $\mathbf{v}$, $\mathbf{p}$, $\mathbf{q}$, $\mathbf s$,  $\mathbf S$, $\mathbf r$, $\mathbf x$, $\mathbf z$, 
and $\pmb \ell$. Although these quantities can be time-varying in practice, their dependence on time is omitted here for simplicity. 

As $\{r_{n}, x_{n}\}_{n \in \cN}$ have relatively small entries, the last terms in both \eqref{eq:sn} and \eqref{eq:vn} are usually ignored, yielding a set of simplified relations, also known as linear distribution flow equations \cite{baran1989optimal}
\begin{subequations}\label{eq:powerflowequationslinear}
    \begin{align}
     s_{n} &=\sum_{i \in \mathcal{C}_{n}} S_{i}-S_{n} \label{eq:snlinear}\\ 
     v_{n} &=v_{\pi_{n}}-2 \operatorname{Re}\left[z_{n}^{\star} S_{n}\right] \label{eq:vnlinear}.
    \end{align}
\end{subequations}
Using the incidence matrix $\bB$, equations \eqref{eq:powerflowequationslinear} can be rewritten in a compact form as follows
\begin{subequations}\label{eq:lindist}
  \begin{align} 
  \mathbf{s} &=\tilde{\bB}^{\top} \mathbf{S} \\
  \tilde{\bB} \mathbf{v} &=2 \operatorname{Re}\left[\mathbf{Z}^{\star} \mathbf{S}\right]-\mathbf{b}_{0} v_{0} 
  \end{align}   
\end{subequations}
where $\mathbf{Z}:= \operatorname{diag}(\mathbf{z})$\footnote{$\operatorname{diag}(\cdot)$ denotes a diagonal matrix whose nonzero entries are given by its argument.}, and $v_0$ is the squared voltage magnitude at the substation. Following~\cite{liang2015smartgrid}, we set $\tilde{\bB}^{-1} \bb_{0}=-\mathbf{1}_{N}$ to be the nodal nominal voltages, hence yielding
\begin{equation}\label{eq:voltlinear}
   \mathbf{v}=2 \mathbf{R} \mathbf{p}+ 2 \mathbf{X} \mathbf{q}+v_{0} \mathbf{1}_{N} 
\end{equation}
where $\mathbf{R}:=\tilde{\bB}^{-1} \operatorname{diag}(\mathbf{r}) {(\tilde{\bB}^{-1})}^{\top} \text { and } \mathbf{X}:= \tilde{\bB} \operatorname{diag}(\mathbf{x}){(\tilde{\bB}^{-1})}^{\top}$. Although the voltage magnitude $v_n$ of bus $n$ in the linearized flow model [cf. \eqref{eq:lindist}-\eqref{eq:voltlinear}] is expressed as a linear function of the voltage magnitude of its parent node $v_{\pi_n}$ (plus the line flow $-2 \operatorname{Re}[z_n^{\star} {S}_n]$), this ignores the nonlinear interactions between  $v_n$ and the voltage magnitudes of its children buses $\{v_i\}_{i\in\mathcal C_n}$ that are present in the nonlinear (exact) branch flow model [cf. \eqref{eq:powerflowequations}]. Indeed, most existing contributions, e.g.,~\cite{probing,liang2017topology}, have relied on this linear approximation model~\eqref{eq:voltlinear} to develop
methods for identifying the grid topology, hence yielding suboptimal performance.
This motivates us to use
the nonlinear distribution flow model to develop a nonlinear grid topology identification algorithm in this next section.


\begin{figure}[t]
\centering
\includegraphics[width =0.45 \textwidth]{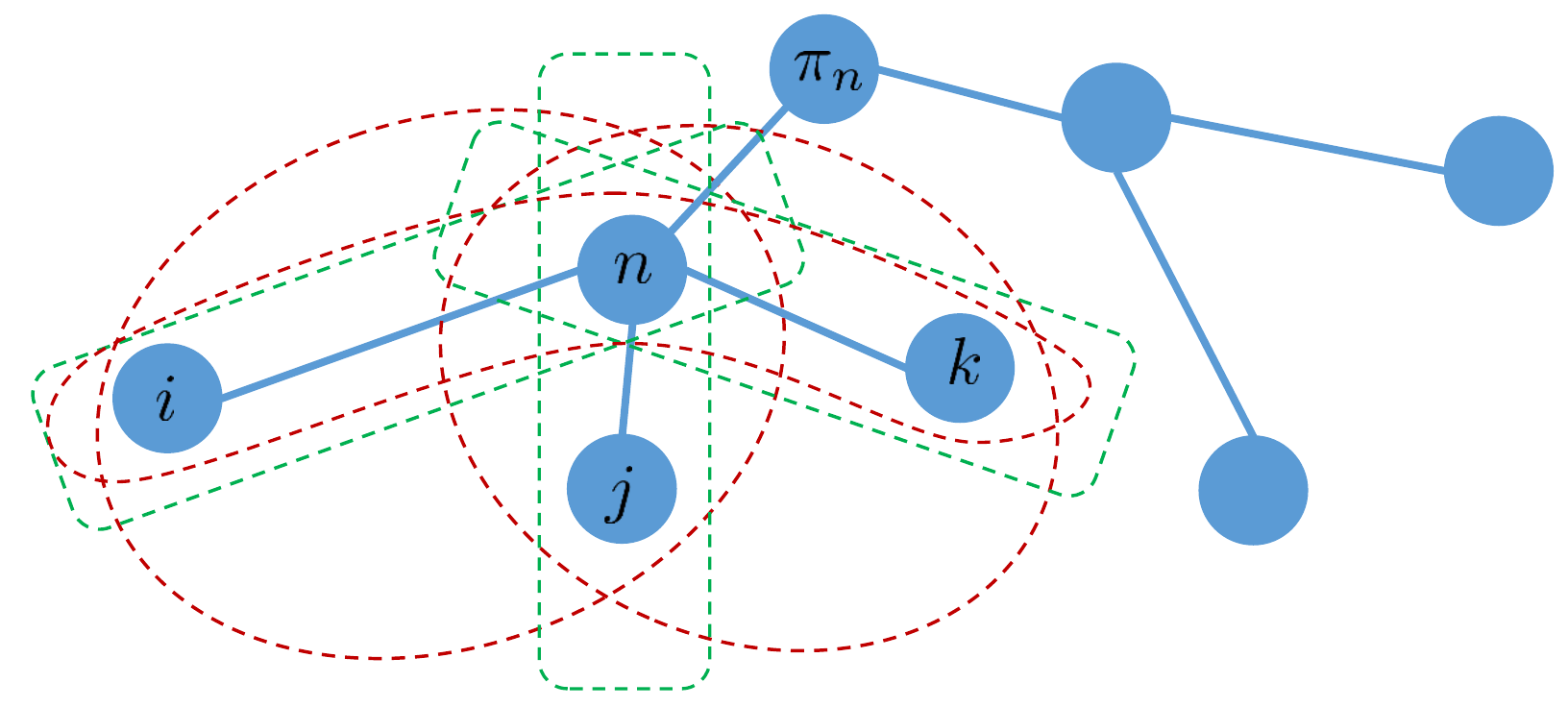}
\caption{Graphical model for triads interactions}
\label{fig:graph}
\end{figure}

\section{Voltage interaction analysis}
\label{sec:prob}
\def\cS{\mathcal{S}}
\def\cC{\mathcal{C}}
Let us elaborate on the dependency of the voltage magnitude $v_n$ of bus $n$ on those of its parent and children buses. Notice from  \eqref{eq:ssn} that $S_n$ is a nonlinear function of $v_{\pi_{n}}$, so we can write $S_n = f_{n}(v_{\pi_n})$ for some nonlinear function $f_n$. Thus, using ~\eqref{eq:snlinear},~\eqref{eq:vnlinear} can be rewritten as follows
\begin{align} \label{eq:nonlinarchild}
v_n =& v_{\pi_n} \!- \!2 \operatorname{Re} \bigg[z_{n}^{\star}  \bigg(\sum\limits_{i: i\in\mathcal{C}_n} \big( \sum_{j: j\in\mathcal{C}_i}f_j(v_i) -s_i + \ell_i z_i \big) -s_n \notag \\
&+ \ell_n z_n\bigg)  \bigg] +\ell_{n}\left|z_{n}\right|^{2} \notag\\
=& v_{\pi_n} \!-\! 2\operatorname{Re}\!\bigg[z_{n}^{\star}\bigg( \sum\limits_{i: i\in\mathcal{C}_n}f_i'(v_i)\!-s_n \!+ \ell_n z_n\bigg) \bigg]\!  \!+\!\ell_{n}\left|z_{n}\right|^{2} 
\end{align}
where we have defined $f_i'(v_i) := \sum_{j: j\in\mathcal{C}_i}f_j(v_i) -s_i + \ell_i z_i $.
Therefore, we can express~\eqref{eq:nonlinarchild} in a compact form as
\begin{equation}\label{eq:nonlinearchildcompact}
	v_n = v_{\pi_n} + g_n(\{v_{i}\}_{i\in\cC_{n}})
\end{equation}
where $g_n(\cdot)$ is a nonlinear function that lumps the second term in~\eqref{eq:nonlinarchild} and highlights the dependency on the children buses of bus $n$. 

Though there are several ways to model the functional $g_n(\cdot)$, e.g., using kernels in \cite{liang2017topology}, 
we are interested in discovering interactions that arise between subsets of buses. Hence, a representation for this function, allowing for interpretability of these higher-order interactions is preferred. In the following, we make use of the recently proposed self-driven graph Volterra models in \cite{coutino2019holink}, to build an explainable model for $g_n$ that elucidates higher-order interactions while capturing the nonlinearity of function $g_n$.

First, let $\cS_{p,l}^{(n)}$ denote a set consisting of $p$ children nodes of bus $n$, with subscript $l$ referring to the $l$-th combination of size $p$, in a lexicographic order, of the children of $n$. For instance,  assuming $i < j < k$ in Fig.~\ref{fig:graph}, we have $\cS_{1,1}^{(n)} = \{i\}$, and $\cS_{2,2}^{(n)} = \{i,k\}$. Therefore, the $l$-th green dashed square and the $l'$-th red dashed ellipse refer to $\cS_{1,l}^{(n)}$ and $\cS_{2,l'}^{(n)}$, respectively. Each of these sets represents possible higher-order interactions among subsets of buses in a distribution network, where only the first-order (i.e., pair-wise) interactions are depicted for illustration purposes.

\def\cV{\mathcal{V}}
Now, let us consider the following model, of order $P$, for approximating the nonlinear function $g_n$ which maps the children voltages to its parent's voltage
\begin{equation}\label{eq:expg}
	g_n(\mathcal{V}_{\cC_n}) \approx \sum\limits_{p=1}^{P}\sum\limits_{l=1}^{L_p}\rho_{p,l}^{(n)}h\Big(\cV_{\cS_{p,l}^{(n)}}\Big)
\end{equation}
where $\cV_{\cS} := \{v_{i}\}_{i\in\cS}$; $L_p$ is the total number of combinations of elements in $\cC_n$ with size $p$; $\rho_{p,l}^{(n)}$ captures the contribution of the set $\cS_{p,l}^{(n)}$ to $v_n$; and $h(\cdot)$ is a nonlinear combining rule for the voltages in its argument. In this work, we consider a multiplicative rule for $h(\cdot)$ for modeling the nonlinear dependency.

Using the expansion~\eqref{eq:expg}, we can rewrite the model for bus voltages in \eqref{eq:nonlinearchildcompact} as follows
\begin{equation}\label{eq:fullmodelvf}
	v_n = v_{\pi_n} + \sum\limits_{p=1}^{P}\sum\limits_{l=1}^{L_p}\rho_{p,l}^{(n)}h\Big(\cV_{\tilde{\cS}_{p,l}^{(n)}}\Big) + \epsilon_n,\quad  \forall\,n\in\cN
\end{equation}
where $\tilde{\cS}_{p,l}^{(n)}$ is defined similarly to ${\cS}_{p,l}^{(n)}$, but with combinations now taken over the superset $\tilde{\cC}_n := \pi_n \cup \cC_n$ instead of $\cC_n$. These new sets allow us to include possible nonlinear dependencies of $v_n$ on $v_{\pi_n}$ too. Here, $\epsilon_n$ captures the modeling error as well as the measurement noise at bus $n$.

The model in~\eqref{eq:fullmodelvf} is akin to the recently proposed self-driven graph Volterra model~\cite{coutino2019holink}, where higher-order interactions over graphs have been used to predict closure events in social networks. Inspired by these results, we found it natural to restrict the sets used in expansion~\eqref{eq:expg} to capture only interactions between two buses and among two buses connected through a central bus. Specifically, we focus on sets defining \emph{edges} and sets defining \emph{$2$-length paths}. That is, for bus $n$, we focus on sets $\{v_i:i\in\cC_n\}$ and sets $\{(v_j,v_i)\,:i,j\in\tilde{\cC}_n\}$. As a consequence, we consider only interactions up to the second order in this paper. Nevertheless, it is worth pointing out that extensions to higher-order interactions, e.g., $k$-length paths, are straightforward.

Since the topology of a distribution grid is unknown a priori, the aforementioned sets cannot be constructed beforehand. Therefore, similar to other network models~\cite{schurgers2002stem}, we propose to fit a sparse model first, enumerating all possible combinations of one and two buses in the network, and subsequently identify such sets (interactions) by their nonzero expansion coefficients.
In this regard, the second-order voltage expansion, considering all relevant groups of bus $n \in \mathcal N$, is given by 
\begin{align}\label{eq:probfinal}
   v_n =& \sum_{i\in\cN_0} \rho_{i}^{(n)} v_{i} 
   +\sum_{i\in \cN_0}\;\sum_{j\in \{k : k \in \cN_0, k \geq i\}} \rho_{i,j}^{(n)} v_iv_j+ \epsilon_n
\end{align}
where $\rho_i^{(n)}$ and $\rho_{i,j}^{(n)}$ are the first- and second-order expansion coefficients relating bus $n$ with the sets $\{i\}$ and $\{i,j\}$, respectively. In~\eqref{eq:probfinal}, we have considered the upper triangular definition of the expansion because functional $h(\cdot)$ is postulated invariant to any permutation of its arguments. The expansion~\eqref{eq:probfinal} can be further expressed in a compact manner as follows
\begin{equation}\label{eq:lastmodelv}
	v_n = \rhob_{n,1}^\top\vb + \rhob_{n,2}^\top(\vb\boxtimes\vb)
\end{equation}
where we have defined $\vb\boxtimes\vb := [v_1^2\; v_1v_2\;\cdots\;v_{N-1}v_N\;v_N^2]^\top$ to be the reduced Kronecker product, while vectors $\rhob_{n,1}$ and $\rhob_{n,2}$ stack up the first- and second-order coefficients $\rho_i^{(n)}$ and $\rho_{i,j}^{(n)}$, in a lexicographic order, respectively.  
The model in~\eqref{eq:lastmodelv} holds for any time slot $t$.

\def\mb{\mathbf{m}}
Now, let us consider a voltage magnitude time-series measurements, collected in $\{\vb(t)\}_{t=1}^{T}$, at time $t =1,\ldots,T$. Denote $\mb(t) := \big[\vb(t)^\top~ (\vb(t)\boxtimes\vb(t))^\top\big]^\top$, and $\bm \theta_n : = \big[\rhob_{n,1}^\top~ \rhob_{n,2}^\top\big]^\top$. Here, $\bm\theta_n$ concatenates the \emph{graph Volterra kernels} for the $n$-th variable. By stacking up different voltage measurements by bus and time into a matrix, that is having elements given by $ V_{n,t} = v_n(t)$, the model for a time-series of voltage magnitude measurements can be represented as follows
\begin{align}
 \Vbo :&=\left[\bm{\theta}_{1}~\bm{\theta}_{2}~\cdots ~ \bm{\theta}_{N}\right]^\top \mathbf{M} + \mathbf{E}\nonumber\\ &=\mathbf{\Theta}^\top \mathbf{M}+\mathbf{E}\label{eq.volSEMMVtheta}
\end{align}
where $\bm M:=[\mb(1),\ldots,\mb(T)]$ and $\mathbf E$ collect the corresponding modeling and measurement errors, respectively.
For interpretability of~\eqref{eq.volSEMMVtheta}, one can rewrite it as
\def \bR{\mathbf R}
\def\bRo{\mathbf R^{(1)}}
\def\bRt{\mathbf R^{(2)}}
\def\bVo{\bV^{(1)}}
\def\bVt{\bV^{(2)}}
\def\bE{\mathbf E}
\begin{equation}\label{eq.volSEMMV}
    \Vbo = \bRo\Vbo + \bRt\Vbt + \bE
\end{equation}
where the $t$-th columns of  $\Vbo$ and $\Vbt$ are $\vb(t)$ and $\vb(t)\boxtimes\vb(t)$, respectively; and, the $n$-th rows of $\bRo$ and $\bRt$ are $\rhob_{n,1}^\top$ and $\rhob_{n,2}^\top$, respectively.

It is worth remarking that, the model~\eqref{eq.volSEMMV} inherits certain desirable characteristics from both SEMs as well as Volterra models. Clearly, it shares the self-driven nature with SEM models, through the first term on the right-hand-side of \eqref{eq.volSEMMV}. It also captures the nonlinear effects as a Volterra series while unveiling the higher-order interactions present in the data. 
This is the reason why this kind of model is known as the self-driven graph Volterra models. Notice that when the graph Volterra coefficients $\bRt$ are set to zero, the model~\eqref{eq.volSEMMV} particularizes to the classical SEM.

\section{Identification of Higher-order Grid Interactions}
\label{sec:solver}

In this section, we start with several assumptions for identifying the graph Volterra coefficients, and formally introduce the optimization problem for finding their values.

For identifying the coefficients of the model~\eqref{eq.volSEMMV}, we make the following assumptions.
\begin{assumption}\label{asu.a1}
The matrix $\bRo$ is a hollow matrix, i.e., $\rho_n^{(n)} = 0,\,\;\forall n \in \mathcal{N}$.
\end{assumption}
\begin{assumption}\label{asu.a2}
The coefficients for the second-order interactions satisfy $\rho^{(n)}_{j,k} = 0$, if $n = j$, or $j = k$, or $n = k$ holds.

\end{assumption}
\begin{assumption}\label{asu.a3}
The graph Volterra coefficients obey $\rho^{(n)}_{j,k} = 0$, if there exists $ \rho_{l}^{(n)} = 0,\, \forall\,l\in\{j,k\}$.
\end{assumption}

While Assumptions A.~\ref{asu.a1} and A.~\ref{asu.a2} can be easily included in an optimization problem, because both involve linear constraints; the last one is a \emph{conditional} constraint. To avoid calling for an alternating minimization, or a mixed-integer program solver, we introduce an auxiliary matrix to enforce Assumption A.~\ref{asu.a3} below when fitting the graph Volterra coefficients.

Let us consider the following matrix
\begin{equation}\label{eq:rmtx}
    \bR_n := \begin{bmatrix}
    \rho_{1}^{(n)} & \rho_{1,1}^{(n)} & \cdots & \rho_{1,N}^{(n)} \\
    \rho_{2}^{(n)} & \rho_{2,1}^{(n)} & \cdots & \rho_{2,N}^{(n)} \\
    \vdots & \vdots & \ddots & \vdots \\
    \rho_{N}^{(n)} & \rho_{N,1}^{(n)} & \cdots & \rho_{N,N}^{(n)} \\
    \end{bmatrix}
\end{equation}
whose first column corresponds to all first-order Volterra kernels of bus $n$, i.e, $\rhob_{n,1}$. By enforcing \emph{row sparsity} in $\bR_n$, $\forall n \in \mathcal N$, we can guarantee that if $\rho_{n,1}(i) = 0$, then  $\rho_{n,2}(i,j)= 0 $, $\forall j\in\mathcal N$. This condition can be effected by using $\ell_{2,1}$-regularization on $\bR_n^\top$.

With relations \eqref{eq.volSEMMVtheta} and \eqref{eq:rmtx}, we consider the following sparsity-aware $\ell_{2,1}$-regularized least-squares for estimating the expansion coefficients
\begin{subequations}
  \label{eq.optProb}
    \begin{align}
    \underset{\{\bm\theta_n\}_{n=1}^{N}} {\min} & ~~ \sum\limits_{n\in \mathcal N}\Vert \vb_n - \mathbf M^\top \bm\theta_n \Vert_2^2 + \lambda \Vert \bm\theta_n \Vert_1 + \mu \Vert \bR_n^\top\Vert_{2,1} \\
    \text{s. to} ~~&~~~  \bm\Theta \in \mathcal{X}_{\rho} .
\end{align}
\end{subequations}
where the dependence of $\thetab_n$ and $\bR_n$ on $\rho_{n,1}(i)$ and $\rho_{n,2}(i,j)$ was omitted for brevity, and the convex set $\mathcal{X}_{\rho}$ signifies the constraints collectively characterized by Assumptions A.~\ref{asu.a1} and A.~\ref{asu.a2}. The optimization problem~\eqref{eq.optProb} is convex, and it can be solved by any off-the-shelf convex programming method. In the following, we examine the performance of our method for identifying the interactions on real datasets.
\begin{figure}
\centering
\psfrag{This paper}{This paper}
\includegraphics[width=0.8\linewidth]{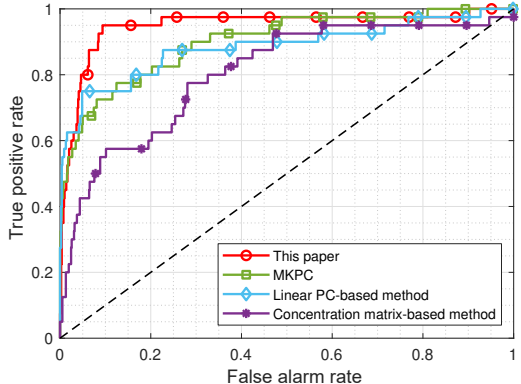}
\caption{ROC curves for topology inference of the SCE 47-bus distribution grid
from voltage magnitude data.}
\label{fig:roc}
\end{figure}

\section{Numerical tests}
\label{sec:test}
For our numerical test, we called for real consumption and solar generation data from the Smart${^*}$ project \cite{barker2012smart} using the SCE $47$-bus distribution grid \cite{farivar2011inverter}, \cite{yang2019tsg}. Using this data, 
voltage squared magnitude measurements $\{\vb(t)\}_{t=1}^{T}$ across $T = 240$ time slots were obtained by solving the AC power flow equations. The voltage magnitudes of the substation bus, $v_0(t)\,\forall\,t \in \{1,\ldots,T\}$, were set to one. After excluding the substation bus as well as the buses connected to their parent buses with zero-impedance lines, we arrived at a radial grid of $41$ buses from which their interactions have to be inferred.

To find the grid topology, we first estimate the graph Volterra kernels in~\eqref{eq.volSEMMVtheta} with the voltage time-series $\{\vb(t)\}_{t=1}^{T}$, by solving~\eqref{eq.optProb} and construct $\bRo$ and $\bRt$ [cf.~\eqref{eq.volSEMMV}]. The grid topology is inferred from the support of $\bRo$ after a point-wise thresholding operation removing non-significant entries. Similarly,  higher-order interactions can be directly retrieved from the support of $\bRt$.

The proposed method was compared with three existing methods, including the multi-kernel based partial correlations (MKPC)- \cite{liang2017topology}, linear PC- \cite{save2013cdc}, and concentration matrix-based~\cite{Deka17} schemes, in terms of edge connectivity recovery performance. This metric has been chosen, since none of the baselines provides information about the higher-order interactions in the grid. 

The empirical receiver operating characteristic (ROC) curves for all  methods are shown in~Fig.~\ref{fig:roc}. 
In addition, their area under the curve (AUC) values are also provided in Table \ref{tab:auc}. These results showcase the merits of exploiting the nonlinear relationships through the exact distribution flow model \eqref{eq:powerflowequations}, relative to 
the linear approximation model \eqref{eq:lindist}-\eqref{eq:voltlinear} that is used in the simulated baselines. Although the MKPC-based scheme captures nonlinearities, these are derived from PCs using the linear approximation model. Finally, the proposed method avoids the computational burden of selecting the proper kernels to capture nonlinearities in the data, while providing insights on the interactions among groups of buses.


\begin{table}[htp]
\vspace{-2mm}
\caption{AUC values for different methods}\vspace{.3cm}
\centering
\renewcommand{\arraystretch}{1.4} 
\begin{tabular}{c c c c }
\hline
\centering This paper & MKPC & Linear PC& Concentration matrix \\
\hline
\centering  \textbf{0.9483} & 0.9008& 0.8836 & 0.8052\\
\hline
\label{tab:auc}
\end{tabular}
\vspace{-9mm}
\end{table}
\section{Conclusions}
\label{sec:conc}
In this work, the problem of unveiling jointly the connectivity and higher-order interactions in a distribution grid was studied. Based on the exact distribution flow model, an expansion relating the voltage magnitude of a set of children buses with their parent's voltage was introduced. This expansion was shown akin to the recently proposed self-driven graph Volterra models devised for higher-order interaction prediction. 
Through this formalism, a topology and higher-order interaction identification method was developed. The merits of considering both the exact grid model as well as the higher-order interactions relative to existing methods were corroborated through numerical tests using real data. This work also opens up interesting directions for future research, including generalizations to (unbalanced) multi-phase distribution grids.

\vspace{.2cm}
\textbf{Acknowledgments.} Q. Yang was  supported in part by NSFC Grants 61522303, 61720106011, 61621063, and the China Scholarship Council. M. Coutino and G. Leus were supported in part by the ASPIRE project 14926 (within the STW OTP program)
financed by the Netherlands Organization for Scientific Research
(NWO); and M. Coutino was also supported by CONACYT. G. Wang and G. B. Giannakis was supported in part by  NSF Grants 1509040,  1711471, and 1901134. E-mail: \{yang6726,\, gangwang,\,georgios\}@umn.edu; \{m.a.coutinominguez,\,g.j.t.leus\}@tudelft.nl.

\bibliographystyle{IEEEtranS}
\bibliography{topology,power}

\end{document}